# Rotating surface solitons

Yaroslav V. Kartashov,[1] Victor A. Vysloukh,[2] and Lluis Torner[1]

[1]*ICFO-Institut de Ciencies Fotoniques, and Universitat Politecnica de Catalunya, Mediterranean Technology Park, 08860 Castelldefels (Barcelona), Spain*

[2]*Departamento de Fisica y Matematicas, Universidad de las Americas – Puebla, Santa Catarina Martir, 72820, Puebla, Mexico*

We introduce a novel type of surface waves that form at the edge of guiding structures consisting of several concentric rings. Such surface waves rotate steadily upon propagation and, in contrast to nonrotating waves, for high rotation frequencies they do not exhibit power thresholds for their existence. There exists an upper limit for the surface wave rotation frequency, which depends on the radius of the outer guiding ring and on its depth.

Under appropriate conditions light can attach to the boundary of materials with different nonlinear properties, which results in surface wave formation. The properties of such waves differ substantially from the properties of solitons in uniform materials. Thus, surface waves usually require threshold power for their formation, which is determined mostly by the difference in refractive indices of materials forming the interface and thus can be made remarkably small in semi-infinite waveguide arrays (or lattices) possessing shallow refractive index modulations [1,2]. Lattice interfaces support not only fundamental [1,2], but also gap surface solitons [3-6]. In two dimensions lattice edges may take on complex shapes and support rich families of nonlinear surface waves [7,8].

In this Letter we introduce a new type of two-dimensional surface waves that exist at the edge of guiding structures consisting of several concentric rings, with a refractive index modulation inside the structure which is periodic along the radial coordinate. The nonzero curvature of the inner and outer rings of such structure drastically alter the properties of surface excitations and affords the existence of rotating surface solitons. Note that solitons in infinite ring-like structures were studied in Bessel lattices [9-12], in photonic bandgap ring fibers [13], and in radially periodic potentials [14,15], while experimental observation of light localization in ring structures was performed in [16]. Here we show that a finite ring structure gives rise to rotating surface waves that, in contrast to nonrotating waves, do not feature power thresholds for their formation if rotation frequency is high enough. The maximal possible rotation frequency increases with the power carried by the surface soliton.

Our analysis is based on the nonlinear Schrödinger equation for the field amplitude $q$ of laser beam propagating along the $\xi$ axis in a focusing cubic medium with a transverse refractive index modulation:

$$i\frac{\partial q}{\partial \xi} = -\frac{1}{2}\Delta_\perp q - |q|^2 q - pR(\eta,\zeta)q, \tag{1}$$

where $\Delta_\perp = \partial^2/\partial\eta^2 + \partial^2/\partial\zeta^2$ is the Laplacian, the transverse $\eta,\zeta$ and the longitudinal $\xi$ coordinates are normalized to the beam width and diffraction length, respectively; $p$ stands for the lattice depth and the function $R(\eta,\zeta)$ describes the transverse refractive index profile. We set $R = \cos^2(\Omega r)$ for $r \leq r_{\text{out}}$ and $R = 0$ for $r > r_{\text{out}}$, where $r^2 = \eta^2 + \zeta^2$ is the radius, $r_{\text{out}} = (2n-1)\pi/2\Omega$ is the outer radius of the guiding structure, $\Omega$ is the radial frequency, and $n$ sets the number of rings. Figure 1(a) shows an example of such a structure with $n = 5$. Such structures can be fabricated permanently [13] or they can be induced optically in suitable crystals [16]. For concreteness, we set $\Omega = 2$.

We look for rotating solitons of Eq. (1) in the form $q = [u(\eta',\zeta') + iv(\eta',\zeta')]\exp(ib\xi)$ (here $u,v$ are the real and imaginary amplitudes and $b$ is the propagation constant) in the rotating coordinate system $\eta' = \eta\cos(\alpha\xi) + \zeta\sin(\alpha\xi)$ and $\zeta' = \zeta\cos(\alpha\xi) - \eta\sin(\alpha\xi)$, where $\alpha$ is the rotation frequency. In these coordinates, the equations for the field components write (we omit the prime coordinates): $\alpha\eta\partial v/\partial\zeta - \alpha\zeta\partial v/\partial\eta + (1/2)\Delta_\perp u + u(u^2 + v^2) + pRu = bu$ and $\alpha\zeta\partial u/\partial\eta - \alpha\eta\partial u/\partial\zeta + (1/2)\Delta_\perp v + v(u^2 + v^2) + pRv = bv$.

The profile of nonrotating solitons $(\alpha = 0)$ residing in the outermost ring is shown in Fig. 1(b). The energy flow $U = \int\int_{-\infty}^{\infty} |q|^2 d\eta d\zeta$ of such a soliton is a nonmonotonic function of $b$ [Fig. 2(a), curve 1] and there exists a lower cutoff on $b$ for the soliton to exist. Low-amplitude solitons tend to penetrate deep into the ring structure [Fig. 1(b)].

Thus, the behavior of nonrotating solitons at such interface is similar to behavior of surface solitons at interfaces of truly periodic lattices [7,8]. This is mostly apparent for $n \to \infty$, when the curvature of the outer ring tends to zero and one gets effectively the interface of quasi-1D lattice. The minimal energy flow and the cutoff $b_{\rm co}$ for $\alpha = 0$ vary only slightly with the number of rings in the structure.

This picture changes drastically as soon as soliton rotation takes place. While for small rotation frequencies $\alpha \sim 0.01$ the dependence $U(b)$ does not change its character, at a certain minimal frequency this curve experiences a substantial deformation so that $U$ vanishes at the cutoff [Fig. 2(a), curve 2]. Even though slowly rotating surface solitons are well localized far from the cutoff [Fig. 1(c)], at $b \to b_{\rm co}$ they show a tendency for expansion not only in the radial direction, but also in the azimuthal one, gradually acquiring an extended azimuthal shape [compare profiles in Figs. 1(d) and 1(b) corresponding to the same values of $b$]. For higher $\alpha$ values the energy flow is a monotonically increasing function of $b$.

Thus, one of the central findings of this Letter is that setting surface solitons into *rotation* at the edge of concentric ring lattice makes them *thresholdless*. The cutoff $b_{\rm co}$ for soliton existence monotonically increases with increase of $\alpha$ [Fig. 3(a)]. The larger the difference $b_{\rm co}(\alpha) - b_{\rm co}(0)$ the stronger the soliton localization along the radial coordinate, even when $b \to b_{\rm co}$. For high rotation frequencies low-power solitons remain well localized in the outer ring of the lattice, but they always strongly expand along the azimuthal direction [Figs. 1(e) and 1(f)]. Concentration of light in the outermost ring even at low powers can be explained by the considerable centrifugal forces acting on rotating states, which result in light expulsion from the central lattice region, and which are counterbalanced by repulsion from its outer edge. At high energy flows rotating surface solitons are well localized.

We found that for a fixed $b$ there exist a maximal possible value of rotation frequency. Thus, for sufficiently large $b > b_{\rm co}(0)$ the energy flow decreases with $\alpha$ and vanishes when $\alpha \to \alpha_{\rm m}$ [Fig. 2(b)]. At $\alpha \to \alpha_{\rm m}$, rotating solitons strongly expand along the azimuthal direction and acquire an extended shape along the ring. Importantly, surface solitons with *higher energies* can rotate with *higher angular frequencies*. In cubic nonlinear media the lattice impacts stronger high-amplitude solitons than low-amplitude ones, in the sense that larger transverse phase tilts are necessary for high-amplitude solitons to move across the lattice. This compensates the higher centrifugal forces acting on such states. The maximal rotation frequency increases with $b$ and goes to zero when $b \to b_{\text{co}}(\alpha=0)$ [Fig. 3(b)]. This is consistent with the fact that linear eigenstates of circular guiding structures modulated in azimuthal direction does not rotate. Nonrotating eigenstates can be represented as a superposition of $\exp(im\phi)$ and $\exp(-im\phi)$ angular harmonics of equal amplitudes. Rotating soliton solutions might emerge from superposition of angular harmonics with unequal amplitudes, acquiring different phase shifts on propagation due to nonlinearity.

Increasing lattice depth $p$ at sufficiently high $\alpha$ does not result in qualitative modification of soliton properties. The maximal rotation frequency for solitons in shallower lattices may be smaller than that in deeper lattices, provided that the propagation constants are selected in such way that at $\alpha=0$ the soliton energy flows in both lattices coincide. This is in agreement with intuitive expectations, since the trapping capability of deeper lattices is higher, hence solitons in such structures may withstand faster rotations. The maximal rotation frequency strongly depends on the number of rings in the lattice (i.e. on $r_{\text{out}}$). For a fixed $b$ the maximal rotation frequency rapidly diminishes with increasing $n$ [Fig. 3(c)], i.e., guiding structures having *smaller outer radius* may support solitons that rotate *faster*. The propagation constant cutoff for rotating solitons growths with $n$.

Integration of Eq. (1) with input conditions $q|_{\xi=0} = (u+iv)(1+\rho)$, where $\rho(\eta,\zeta)$ stands for broadband noise with variance $\sigma_{\text{noise}}^2$, show that rotating solitons are stable for both small (in this case $b$ should not be too close to $b_{\text{co}}$, where derivative $dU/db$ becomes negative) and high rotation frequencies (in this case solitons are stable in the entire existence domain in $b$). Figure 4 shows stable rotation of well-localized surface solitons with moderate frequencies and solitons strongly elongated in the azimuthal direction with high rotation frequencies close to $\alpha_{\text{m}}$ for corresponding $b$ and $n$ values. We did not observe any radiative losses even for rapidly rotating solitons after propagation over considerable distances. Rotating surface solitons discussed here may arise upon development of azimuthal modulational instabilities in circular structures with multiple guiding rings [17].

Summarizing, we introduced a new type of rotating surface solitons that exist at the edge of guiding structures consisting of several concentric rings. Such surface solitons may be thresholdless for high enough rotation frequencies. High-amplitude surface waves were found to rotate faster than their low-amplitude counterparts.

# References with titles

# References without titles

# Figure captions

Figure 1.  (a) Ring guiding structure corresponding to $n=5$. Field modulus distributions for surface solitons at (b) $b=6.45$, $\alpha=0$, (c) $b=7$, $\alpha=0.1$, (d) $b=6.45$, $\alpha=0.1$, (e) $b=9.3$, $\alpha=0.4$, and (f) $b=7$, $\alpha=0.4$. (c) In all plots $p=10$ and $\Omega=2$. White circles indicate the center of the outer ring of guiding structure. In (b)-(e) $n=5$, while in (f) $n=3$.

Figure 2.  (a) Energy flow versus propagation constant for $\alpha=0$ (curve 1), $0.1$ (curve 2), and $0.15$ (curve 3). Points marked by circles correspond to soliton profiles shown in Figs. 1(b)-1(d). (b) Energy flow versus rotation frequency for different propagation constants. In all cases $p=10$, $\Omega=2$, and $n=5$.

Figure 3.  (a) Propagation constant cutoff versus rotation frequency at $n=5$. (b) Maximal rotation frequency versus propagation constant at $n=5$. (c) Maximal rotation frequency versus number of rings in guiding structure at $b=9$. In all cases $p=10$ and $\Omega=2$.

Figure 4.  Snapshot images showing stable rotation of surface solitons at (a) $\alpha=0.1$, $n=5$ (b) $\alpha=0.29$, $n=5$ (c) $\alpha=0.1$, $n=3$ and (d) $\alpha=0.51$, $n=3$. In all cases input solitons correspond to $b=8$, $p=10$, and $\Omega=2$. The right outermost image in each plot shows input field modulus distribution in the presence of white noise with variance $\sigma_{\text{noise}}^2=0.01$, while two other images superimposed onto input distribution are taken at proper distances exceeding $\xi=500$. White circles indicate the center of the outer ring of guiding structure.

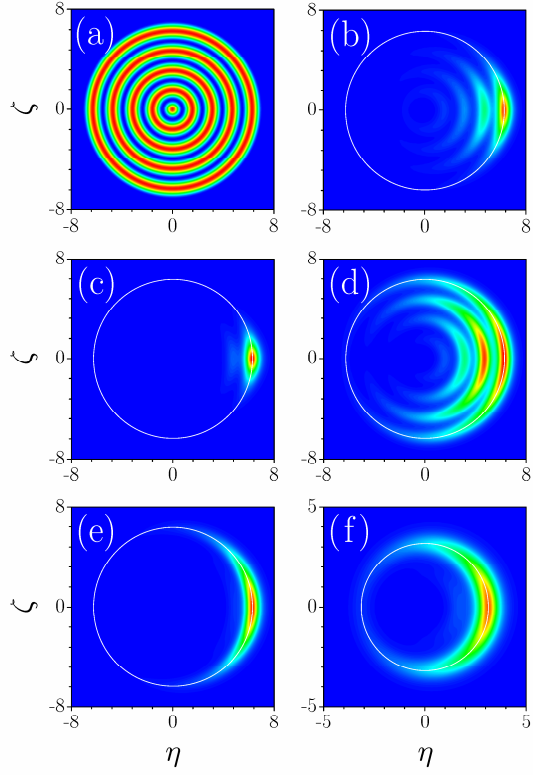

Figure 1. (a) Ring guiding structure corresponding to $n=5$. Field modulus distributions for surface solitons at (b) $b=6.45$, $\alpha=0$, (c) $b=7$, $\alpha=0.1$, (d) $b=6.45$, $\alpha=0.1$, (e) $b=9.3$, $\alpha=0.4$, and (f) $b=7$, $\alpha=0.4$. (c) In all plots $p=10$ and $\Omega=2$. White circles indicate the center of the outer ring of guiding structure. In (b)-(e) $n=5$, while in (f) $n=3$.

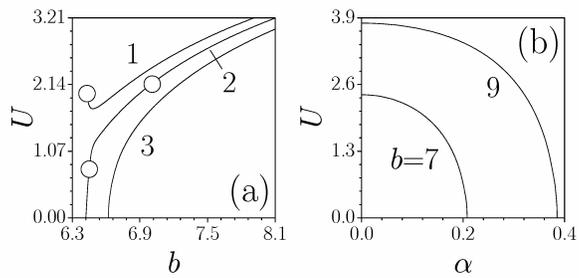

Figure 2. (a) Energy flow versus propagation constant for $\alpha = 0$ (curve 1), $0.1$ (curve 2), and $0.15$ (curve 3). Points marked by circles correspond to soliton profiles shown in Figs. 1(b)-1(d). (b) Energy flow versus rotation frequency for different propagation constants. In all cases $p = 10$, $\Omega = 2$, and $n = 5$.

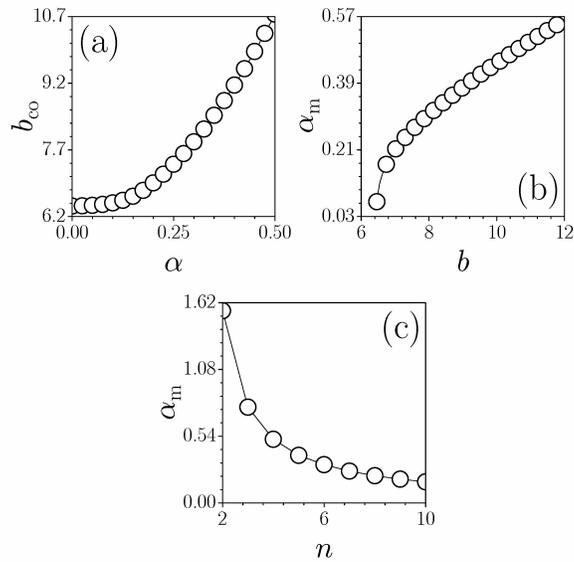

Figure 3. (a) Propagation constant cutoff versus rotation frequency at $n=5$. (b) Maximal rotation frequency versus propagation constant at $n=5$. (c) Maximal rotation frequency versus number of rings in guiding structure at $b=9$. In all cases $p=10$ and $\Omega=2$.

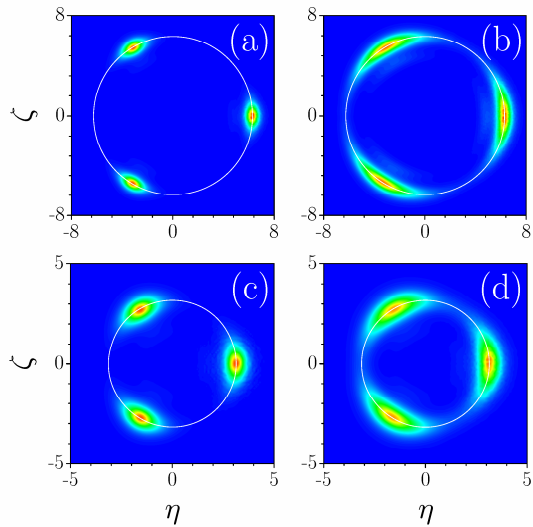

Figure 4. Snapshot images showing stable rotation of surface solitons at (a) $\alpha = 0.1$, $n = 5$ (b) $\alpha = 0.29$, $n = 5$ (c) $\alpha = 0.1$, $n = 3$ and (d) $\alpha = 0.51$, $n = 3$. In all cases input solitons correspond to $b = 8$, $p = 10$, and $\Omega = 2$. The right outermost image in each plot shows input field modulus distribution in the presence of white noise with variance $\sigma_{\text{noise}}^2 = 0.01$, while two other images superimposed onto input distribution are taken at proper distances exceeding $\xi = 500$. White circles indicate the center of the outer ring of guiding structure.